\documentclass[twocolumn,showpacs,preprintnumbers,amsmath,amssymb,nofootinbib]{revtex4}

\usepackage{graphicx}
\usepackage{dcolumn}
\usepackage{bm}
\usepackage{amssymb}
\usepackage{amsmath}
\usepackage{amsthm}

\begin{document}


\title{Nuclear and electronic contributions to Coulomb correction 
for Moli\`{e}re screening angle}

\author{M.V.~Bondarenco}
 \email{bon@kipt.kharkov.ua}
 \affiliation{%
NSC Kharkov Institute of Physics and Technology, 1 Academic St.,
61108 Kharkov, Ukraine }

\date{\today}

\begin{abstract}

The Coulomb correction (difference from the 1st Born approximation) 
to the Moli\`{e}re screening angle in multiple Coulomb scattering theory is evaluated with the allowance for inelastic contribution. 
The controversy between dominance of close- or remote-collision contributions to Coulomb correction is discussed. 
For scattering centres represented by a Coulomb potential with a generic (not necessarily spherically symmetric) screening function, the Coulomb correction is proven to be screening-independent, 
by virtue of the eikonal phase cancellation in regions distant from the Coulomb singularity. 
Treating the atom 
as an assembly of pointlike electrons and the nucleus, 
and summing the scattering probability over all the final atom states, 
it is shown that besides the Coulomb correction due to close encounters of the incident charged particle with atomic nuclei, there are similar corrections due to close encounters with atomic electrons 
(an analog of Bloch correction). 
For low $Z\neq1$ the latter contribution can reach $\sim 25\%$, 
but its observation is partly obscured by multiple scattering effects.


\end{abstract}





\maketitle

\section{Introduction}

The term ``Coulomb correction'' in physics of high-energy atomic collisions conventionally designates deviation from the 1st Born approximation for certain observables, 
for which it depends solely on the Coulomb parameter -- the product of nuclear charges of the colliding particles and their reciprocal collision velocity, but not on the screening function. 
In 1930-1950-ies, corrections of that kind were independently discovered for
ionization energy loss \cite{Bloch,Lindhard-Sorensen,Sorensen,Khodyrev,Matveev-Makarov-Gusarevich}, 
multiple Coulomb scattering \cite{Moliere,BK-LPM,Kuraev-Tarasov-Voskr}, 
bremsstrahlung \cite{Davies-Bethe-Maximon,Olsen-Maximon-Wergeland,Lee-Milstein-Strakhovenko-Schwartz}, 
and electron-positron pair production 
\cite{Olsen,Lee-Milstein,Ivanov-Schiller-Serbo}. 
For different processes, different formalisms were applied, 
such as partial wave expansion \cite{Bloch,Lindhard-Sorensen,Sorensen}, 
eikonal approximation \cite{Moliere,BK-LPM,Lee-Milstein-Strakhovenko-Schwartz,Lee-Milstein,Kuraev-Tarasov-Voskr,Matveev-Makarov-Gusarevich}, 
Furry-Sommerfeld-Maue wave functions \cite{Davies-Bethe-Maximon,Olsen,Ivanov-Schiller-Serbo,Khodyrev}, 
so, the final results were sometimes presented in different forms, as well. 

In pioneering work \cite{Bloch} a correction
\begin{equation}\label{Bloch}
\Delta L_{\text{Bloch}}=-f\left(\frac{Z_1e^2}{\hbar v}\right),
\end{equation}
to stopping number $L$ entering Bethe-Bloch formula
\begin{equation}\label{Bethe-Bloch}
\frac{dE}{dx}=\frac{4\pi n_a Z}{m_e}\left(\frac{Z_1 e^2}{v}\right)^2L
\end{equation}
was expressed in terms of  function
\begin{equation}\label{f-def}
f\left(s\right)=\mathfrak{Re}\,\psi\left(1+is\right)-\psi(1),\qquad f(-s)=f(s),
\end{equation}
where $\psi(s)=\Gamma'(s)/\Gamma(s)$ is digamma function, 
$Z_1$ and $Z$ are the incident particle and the target atom nucleus charges in units of the proton charge $e$,
$e^2/\hbar=c/137$, 
$n_a$ the substance density, and $m_e$ the electron mass.

In \cite{Davies-Bethe-Maximon}, evaluating spectra of bremsstrahlung from an ultrarelativistic electron in a bare Coulomb field, 
and $e^+ e^-$ pair production in such a field, 
a nonperturbative correction to the large logarithm was obtained in form of  function (\ref{f-def}) having argument $\frac{Ze^2}{\hbar c}$, 
with $Ze$ being the charge of the target nucleus.

Instead, in \cite{Moliere}, 
where a theory of multiple scattering on screened Coulomb potentials of atoms in matter was developed, 
the universal dependence of the Moli\`{e}re screening angle [defined below in Eqs. (\ref{qa-lim-int}), (\ref{chia=qap})] on Coulomb parameter  $\frac{Z_1Ze^2}{\hbar v}$, 
was presented in form of an algebraic interpolation 
\begin{equation}\label{Moliere-sqrt}
\frac{\chi_a\left(\frac{Z_1Ze^2}{\hbar v}\right)}{\chi_a(0)}
=\sqrt{1+\frac{3.76}{1.13}\left(\frac{Z_1Ze^2}{\hbar v}\right)^2}.
\end{equation}
Later, it was eventually realized that those corrections are essentially of the same physical origin, 
and in \cite{Kuraev-Tarasov-Voskr}
it was argued (see also \cite{Lee-Milstein,Scott}, Sec. VI F) that (\ref{Moliere-sqrt}) should more precisely be expressed via exponentiated function (\ref{f-def}):
\begin{equation}\label{q_a-propto-e^f}
\frac{\chi_a\left(\frac{Z_1Ze^2}{\hbar v}\right)}{\chi_a(0)}
=e^{f\left(\frac{Z_1Ze^2}{\hbar v}\right)}.
\end{equation}

The celebrated simple results involving function (\ref{f-def}), however, 
evoked physical controversies for a long time. 
The interpretation is obscured by encountered multiple integrals with singular integrands regularized by subtraction terms. 
In application to Bloch correction [Eqs. (\ref{Bloch})--(\ref{f-def})] 
a paradox was pointed out in \cite{Lindhard-Sorensen}: 
If, as implied by formula (\ref{Bloch}), 
the correction does not depend on atomic electron binding, 
it must stem from close collisions; 
but the close-collision contribution to the differential cross section is described by Rutherford formula, 
which coincides with its 1st Born approximation, 
whence the non-perturbative correction should be absent at all. 
To resolve this paradox, 
\cite{Sorensen} investigated dependence of the energy loss on the angle of deflection of the incident ion, 
and found the Bloch correction to correspond to small rather than large deflection angles. 
But given the reciprocal relationship between the momentum transfer and impact parameter, 
that seems to contradict the original assumption of its origin from small impact parameters. 
Disturbing may also be the fact that an analog of Coulomb correction in straggling (including an additional weighting factor proportional to the deflection angle squared) is absent, 
whereas close collisions usually give large fluctuations, 
as is, e. g., in Landau theory \cite{Landau}.
Other researchers \cite{Matveev-Makarov-Gusarevich} thus searched for non-Bloch nonperturbative contributions stemming from large impact parameters.


A similar discussion was held for $e^+ e^-$ pair production in collisions of bare heavy ions \cite{Lee-Milstein,Baltz}.

For Coulomb corrections in elastic scattering, 
e. g., in Moli\`{e}re's theory of multiple small-angle scattering in amorphous matter, 
or in theories of bremsstrahlung and $e^+ e^-$ pair production on atoms based on Furry-Sommerfeld-Maue wavefunctions, 
the situation is largely the same, but the main concern is about accuracy, 
with which the Coulomb correction is independent of the Coulomb field screening. 
To simplify analytic evaluation of the encountered multiple integrals, 
\emph{presuming} the dominant contribution to come from small impact parameters, 
the screening is usually merely neglected from the outset. 
The discussion quoted in the previous paragraph can cast doubts upon legitimacy of such a procedure, 
but for any specific choice of the scattering function, 
the difference between (\ref{f-def}) and the prediction of eikonal approximation for scattering in a screened potential can be numerically shown to be compatible with zero. 
Therefore, result (\ref{f-def}) may actually be exact, 
though a general proof of that statement is lacking.

In practical applications of fast charged particle passage through solids, 
it is essential yet that albeit scattering is predominantly elastic, 
there is also an inelastic contribution, 
which becomes relatively sizable at low $Z$. 
To describe inelastic processes, electrons must be treated as pointlike particles rather than a mean charge distribution. 
Therewith, modifications of the inelastic screening angle arise not only at the Born level (as is assumed in \cite{Fano}), 
because collisions with electrons can be nonperturbative, as well. 
Even though the electron charge is lower than that of the nucleus 
(for $Z>1$), 
the Coulomb parameter includes also a factor $Z_1/v$, which can be sufficiently large.
Then, if Bloch correction (\ref{Bloch}), 
differing from the elastic scattering contribution (\ref{q_a-propto-e^f}) by the absence of $Z$ factor in the argument of $f$, 
exists in energy loss \cite{Bloch-experim}, 
in turn being related to the transport scattering cross section \cite{Lindhard-Sorensen}, 
Bloch correction must as well enter the energy-inclusive theory of multiple Coulomb scattering. 
Its incorporation is also desirable for unified theory of multiple scattering and ionization energy loss \cite{Bond-correl} extended to particles heavier than electrons.


In the present article, after shedding some light on the controversy between small- and large-distance contributions (Sec. \ref{sec:prelim}), 
we employ  eikonal approximation to extend the derivation of Coulomb correction to scattering fields of non-spherical configuration (Sec. \ref{sec:2d}). 
The developed integration technique is further used to calculate inelastic scattering on an atom as a few-body system, including pointlike atomic electrons (Sec. \ref{sec:inel-scat}). 
Evaluating the corresponding eikonal scattering amplitude, 
and summing its square over final and averaging over all the the initial state, 
we are led to a Coulomb correction comprised of contributions described by function (\ref{f-def}), 
but with different ($Z$-dependent) weighting factors and arguments, 
corresponding to scattering on atomic nucleus, as well as on individual atomic electrons. 
Estimates for its experimental verification are provided in Sec. \ref{sec:pract}. 






\section{Preliminary considerations}\label{sec:prelim}

Before turning to formal evaluation of integrals defining the Coulomb correction, 
it will be helpful to elucidate its key features by simple examples. 

\subsection{Application of closure identity. 
Absence of higher-Born corrections for finite $\left\langle q^2\right\rangle$}\label{subsec:mean-q2}

Generally, Coulomb corrections arise for quantities related to the transport cross-section or, equivalently, the mean square momentum transfer
\begin{equation}\label{mean-q2}
\left\langle q^2\right\rangle\propto \int d^2q q^2 \frac{d\sigma}{d^2q},
\end{equation}
where ${d\sigma}/{d^2q}$ is the differential cross section of fast particle small-angle deflection with (predominantly transverse) momentum transfer $\bm q$.
Due to the weighting factor $q^2$, the relative role of large momentum transfers, i.e., small transverse distances, is enhanced, 
whereas the dependence on the screening function diminishes, 
but a priori needs not be extinguished entirely.

In the simplest case of scattering of a particle with charge $Z_1e$ and velocity $v$ by a fixed electrostatic potential $\varphi(\bm r)$, 
the differential scattering cross section in the leading order of high-energy expansion (relativistic or nonrelativistic) can be expressed in the eikonal approximation \cite{Moliere,Scott,Glauber,eik-relat}
\begin{equation}\label{dsigmad2q-eik}
\frac{d\sigma}{d^2q}=\left|a\right|^2,
\end{equation}
where
\begin{equation}\label{a-def}
a(\bm q)=\frac1{2\pi i\hbar}\int d^2b
e^{\frac{i}{\hbar}\bm{q}\cdot\bm{b}}\left[1-e^{\frac{i}{\hbar}\chi_0(\bm{b})}\right]
\end{equation}
is the eikonal scattering amplitude, and
\begin{equation}\label{chi0-def}
\chi_0(\bm{b})=-\frac{Z_1 e}{v}\int_{-\infty}^{\infty}dz \varphi(z,\bm{b})
\end{equation}
the eikonal phase depending on the impact parameter $\bm b$.
Physically, the longitudinal coordinate $z$-integral in (\ref{chi0-def}) runs over the particle trajectory, 
which at high energy is nearly straight within the scattering field region. 
Beyond this region, the distorted wave function expands into a superposition of deflected plane waves, 
whence contributions from different trajectories interfere 
(similarly to Fraunhofer diffraction on a finite obstacle), 
as taken into account by the integral over the impact parameters in (\ref{a-def}).
When the interference is small, the stationary phase approximation in the impact parameter plane applies \cite{Moliere,Scott,Glauber,eik-relat}, 
leading to the classical picture [Eq. (\ref{q=gradchi0}) below].

An expedient transformation of the integrand in (\ref{mean-q2}) is
\begin{equation}\label{q2-under-a2}
q^2 \frac{d\sigma}{d^2q}=\left|\bm q a\right|^2.
\end{equation}
Carrying factor $\bm{q}$ under the integral sign, 
transforming it into a gradient acting on the plane wave
\[
\bm{q}e^{\frac{i\bm{q}\cdot\bm{b}}{\hbar}}=\frac{\hbar}{i} \frac{\partial}{\partial\bm{b}}
 e^{\frac{i\bm{q}\cdot\bm{b}}{\hbar}},
\]
and switching its action onto the eikonal phase factor by partial integration, 
one eliminates $\bm{q}$ everywhere except the plane wave factor:
\begin{equation}\label{qa}
\bm q a
=\frac{1}{2\pi}\int d^2b
e^{\frac{i}{\hbar}\bm{q}\cdot\bm{b}}\frac{\partial}{\partial\bm{b}}\left[1-e^{\frac{i}{\hbar}\chi_0(\bm{b})}\right].
\end{equation}
This brings the integral to an ordinary Fourier form.
The benefit is that if in Eqs. (\ref{mean-q2}), (\ref{q2-under-a2}), (\ref{qa}) the integral is finite when extended over the entire $\bm{q}$ plane, 
its evaluation becomes nearly trivial making use of the closure identity:
\begin{eqnarray}\label{ms=msBorn}
\int d^2q q^2 \frac{d\sigma}{d^2q}
=\int d^2b
\left|\hbar\frac{\partial}{\partial\bm{b}}e^{\frac{i}{\hbar}\chi_0(\bm{b})}\right|^2\nonumber\\
 =\int d^2b
\left|\frac{\partial}{\partial\bm{b}}\chi_0(\bm{b})\right|^2
=\int d^2q q^2 \frac{d\sigma_1}{d^2q},
\end{eqnarray}
where $\frac{d\sigma_1}{d^2q}$ is the first Born approximation for $\frac{d\sigma}{d^2q}$, 
obtained by letting $\chi_0\to0$, $e^{\frac{i}{\hbar}\chi_0}\to1+\frac{i}{\hbar}\chi_0$.
The exact result thus appears to be equal to the first Born approximation [as well as to the classical result, since $\frac{\partial}{\partial\bm{b}}\chi_0(\bm{b})$ is the classical momentum transfer (\ref{q=gradchi0})] \cite{Artru}. 
There would then arise no Lindhard-Sorensen paradox.

However, for a screened Coulomb potential, the differential cross section at large $q$ has the Rutherford asymptotics
\begin{equation}\label{Ruth-def}
\frac{d\sigma}{d^2q}\underset{q\to\infty}\simeq \frac{d\sigma_R}{d^2q}, \qquad 
\frac{d\sigma_R}{d^2q}\propto q^{-4},
\end{equation}
wherewith integral (\ref{mean-q2}) logarithmically diverges at large $q$, 
and relation (\ref{ms=msBorn}) becomes meaningless. 
Furthermore, even the difference
\begin{equation}\label{construct-the-difference}
\int d^2q q^2 \left(\frac{d\sigma}{d^2q}-\frac{d\sigma_1}{d^2q}\right),
\end{equation}
which is finite due to mutual cancellation of the Rutherford ``tails'' in the integrand, may be non-zero.
 
Application of the closure identity to eikonal scattering in potentials with Coulomb singularity thus needs caution, but if applied properly, 
it can be a valuable tool. 
It may also be noted that in the high-energy case the impact parameter representation can be more convenient than partial wave expansion, 
since the latter relates phase shifts $\delta_l$ and $\delta_{l+1}$ corresponding to different angular momenta $l$. 
(At large $l$, typical at high energy, though, this difference is small.)


\subsection{Large-$q$ regularization for Coulomb scattering}

Self-consistent physical theories should only operate with finite quantities. 
If a divergence arises in some approximation, it needs to be cured by 
a more accurate approximation, furnishing effective regularization. 
For logarithmically divergent integrals, to which (\ref{mean-q2}) belongs, 
any regularization is equivalent to a sharp cutoff. 
As a representative, regularization scheme-independent part of the next-to-leading logarithmic contribution, 
one can consider 
\begin{equation}\label{qa-lim-int}
\ln q_a(Z_1/v)
=\underset{q_R\to\infty}\lim\left(\ln q_R-\int_0^{q_R} \frac{dq}{q}\frac{d\sigma}{d\sigma_R}\right)-\frac12,
\end{equation}
where $d\sigma_{R}$ is given by (\ref{Ruth-def}). 
Notation $q_a$ and term $1/2$ in the rhs of (\ref{qa-lim-int}) is introduced so that rescaling
\begin{equation}\label{chia=qap}
\chi_a=q_a/p
\end{equation}
by the large longitudinal (nearly conserved) momentum $p$ yields Moli\`{e}re's screening angle $\chi_a$ \cite{Moliere,Bethe}.\footnote{Another commonly used notation, simply related to $\chi_a$, is 
\begin{equation}\label{chi'a-def}
\chi'_a=e^{\gamma_{\text{E}}-1/2}\chi_a
\end{equation}
\cite{Bethe}. 
Since this is a proportionality relationship, while we will study mainly relative deviations, they are the same for both definitions.}
A reminder of how quantity (\ref{chia=qap}) arises in the multiple Coiulomb scattering theory is given in Appendix \ref{app:how-qa-arises}.

Eq. (\ref{qa-lim-int}) can be recast as
\begin{eqnarray}\label{}
\ln q_a+\frac12
=\underset{q_R\to\infty}\lim\left(\ln q_R-\int_0^{q_R} \frac{dq}{q}\frac{d\sigma_1}{d\sigma_R}\right)\nonumber\\
+\int_0^{\infty} \frac{dq}{q}\frac{d\sigma_1-d\sigma}{d\sigma_R}.
\end{eqnarray}
Identification of the limit in the right-hand side with $\frac12+\ln q_a(0)$ leads to relationship
\begin{equation}\label{ln-qa=intdsigma-dsigma1}
\ln \frac{q_a(Z_1/v)}{q_a(0)}
=\int_0^{\infty} \frac{dq}{q}\frac{d\sigma_1-d\sigma}{d\sigma_R}.
\end{equation}
Since $d\sigma_R\propto dq/q^3$, the right-hand side of (\ref{ln-qa=intdsigma-dsigma1}) is proportional to (\ref{construct-the-difference}). 
If it was zero, that would imply ${q_a(Z_1/v)}\equiv{q_a}(0)$. 
But numerical calculation based on equation (\ref{qa-lim-int}) and (\ref{dsigmad2q-eik})--(\ref{chi0-def}) gives a positive function rising indefinitely with the increase of $Z_1/v$, 
and providing the evidence for existence of the Coulomb correction.




\subsection{Classical limit. 
Coulomb correction as a manifestation of global $q$-scaling}

The simplest interpretation of the Coulomb correction can be given in high-energy classical mechanics. 
In that case, $\bm q$ is completely determined by the particle impact parameter 
$\bm b$:
\begin{equation}\label{q=gradchi0}
\bm q(\bm b)=\frac{\partial}{\partial\bm b}\chi_0,
\end{equation}
where $\chi_0(b)$ is defined by Eq. (\ref{chi0-def}). 
Assuming the Coulomb field [characterized by singularity $\varphi(r)\underset{r\to0}\sim\frac{Ze}{r}$, with the nucleus charge $Z$] to be monotonically and spherically symmetrically screened at finite $r$, 
the scattering indicatrix may be expressed as
\begin{equation}\label{q-int-ne}
q(b)=\frac{2Z_1 Z e^2}{vb} S_1(b),
\end{equation}
where $S(b)$ is also a momotonically decreasing function with extremities 
\[
S_1(0)=1, \qquad S_1(\infty)=0. 
\]
Inverse function $b(q)$ is then single-valued, too, whence the classical differential cross section derives from (\ref{q-int-ne}) as a Jacobian
\begin{equation}\label{dsigmacl=dbdq}
\frac{d\sigma_{cl}}{d^2q}
=2\pi b\left|\frac{db}{dq}\right|. 
\end{equation}

Setting in (\ref{q-int-ne}), (\ref{dsigmacl=dbdq}) $S_1\equiv1$, one obtains the familiar formula for the corresponding pure Rutherford scattering differential cross section:
\begin{equation}\label{Ruth-coher}
\frac{d\sigma_R}{dq}=\frac{2\pi}{q^3}\left(\frac{2Z_1 Z e^2}{v}\right)^2.
\end{equation}
Ratio $\frac{dq}{q}\frac{d\sigma}{d\sigma_R}$ entering (\ref{qa-lim-int}) then expresses through $S_1^2(b)$ and the Coulomb singularity factor as
\[
\frac{dq}{q}\frac{d\sigma_{cl}}{d\sigma_R}=\frac{1}{2\pi}\left(\frac{vq}{2Z_1 Z e^2}\right)^2d\sigma_{cl}
=\frac{S_1^2d\sigma_{cl}}{2\pi b^2}.
\]

Changing in (\ref{qa-lim-int}) from integration over $q$ to integration over $b$, i.e., writing $d\sigma_{cl}=2\pi bdb$, we obtain:
\begin{subequations}
\begin{equation}\label{ln-qa=ln[-]}
\ln q_a(Z_1/v)+\frac12=
\lim_{q_R\to\infty}\left[\ln q_R
-\int_{\frac{2Z_1Ze^2}{v q_R}}^{\infty}\frac{db}{b} S_1^2(b) \right],
\end{equation}
Introducing 
$b_R=b_R\left(Z_1Ze^2/\hbar v, q_R\right)=\frac{2Z_1Ze^2}{v q_R}$, and transforming limit $q_R\to\infty$ to $b_R\to0$,
we can equivalently present $q_a$ as a sum of two terms
\begin{eqnarray}\label{ln-qa=ln-alpha+Ca}
\ln q_a(Z_1/v)+\frac12=
\ln\frac{2Z_1Ze^2}{v}\qquad\qquad\qquad \nonumber\\
+ 
\lim_{b_R\to0}\left[\ln\frac{1}{b_R}
-\int_{b_R}^{\infty}\frac{db}{b} S_1^2(b) \right].
\end{eqnarray}
\end{subequations}
The second term (the limit of a difference) in the right-hand side of (\ref{ln-qa=ln-alpha+Ca}) depends only on screening (i. e., on $Z$), 
but not on $Z_1/v$. 
Therefore, the dependence of $q_a$ on the latter parameter is a plain proportionality: 
\begin{equation}\label{qa-propto-Z1v}
q_a(Z_1/v)\propto\frac{Z_1}{v}. 
\end{equation}
That merely reflects the property of classical scaling, 
when the imparted momentum is proportional to the strength of the force acting on the particle, i.e., to its charge, 
and to the action time, which is reciprocal to $v$. 

Comparing Eq. (\ref{ln-qa=ln-alpha+Ca}) with Eq. (\ref{q_a-propto-e^f}), 
which in the classical limit $s\to\infty$ simplifies to 
\begin{equation}\label{fs->infty}
f(s)\underset{s\to\infty}\simeq \ln s+\gamma_{\text{E}},
\end{equation} 
i.e., 
\begin{equation}\label{qa-classical-limit}
q_a(Z_1/v)\underset{Z_1e^2/\hbar v\to\infty}
\simeq \frac{Z_1Ze^2}{\hbar v} e^{\gamma_{E}} q_a(0),
\end{equation}
one infers
\begin{equation}\label{qa-lim-bto0}
\ln \frac{q_a(0)}{2\hbar}+\frac12+\gamma_{\text{E}}=\lim_{b_R\to0}\left[
\ln\frac{1}{b_R}
-\int_{b_R}^{\infty}\frac{db}{b} S_1^2(b) \right].
\end{equation}
While factor $Z_1/v$ here complies with Eq. (\ref{qa-propto-Z1v}), 
the presence of Euler's constant $\gamma_{\text{E}}=-\psi(1)=0.577$ here, as well as factor $e^{\gamma_{\text{E}}}$ in Eq.~(\ref{qa-classical-limit}) looks nontrivial.
They are not of the classical origin, and can be attributed to evolution of the differential cross section at moderate Coulomb parameters.

Now from Eq. (\ref{ln-qa=ln[-]}), 
where the $b$-cutoff at a fixed $q_R$ depends on $Z_1/v$, 
whereas distribution $S_1^2(b)/b$ does not, 
it follows that the Coulomb correction stems from asymptotically small $b$, 
while moderate $b$ always give the same contribution. 

\begin{figure}
\includegraphics{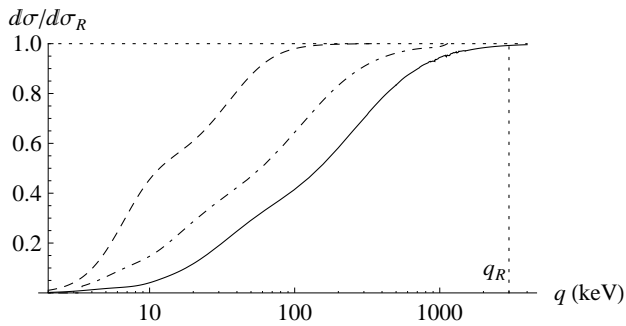}
 \caption{\label{fig:q-shift} Differential cross-section (\ref{dsigmad2q-eik}), (\ref{a-def}) of eikonal scattering in the mean potential of a carbon atom ($Z=6$). 
Plotted are ratios $d\sigma/d\sigma_R$ vs $\log q$; 
the area under such curves is related to $q_a$. 
Dashed, perturbative scattering regime ($Z_1Ze^2/\hbar v=0.1$).
Dot-dashed, $Z_1Ze^2/\hbar v=2$.
Solid, $Z_1Ze^2/\hbar v=5$. 
The latter two curves pertain to classical scattering regime, 
thus differing merely by a translation in $\log q$.
}
\end{figure}

In the momentum transfer representation (differing from $b$-representation merely by a change of the integration variable), however, 
the situation looks just the opposite. 
If, as is customarily done \cite{WilliamsPR1939,Moliere,Sorensen},
$d\sigma/d\sigma_R$ is plotted vs $\log q$ (see Fig. \ref{fig:q-shift}), 
the area under such a curve is related to $q_a$ by 
Eq. (\ref{qa-lim-int}).
The area between two such curves 
corresponding to different Coulomb parameters (e.g., the dot-dashed and solid curves in Fig. \ref{fig:q-shift}) equals to the increment of the Coulomb correction. 
But evidently, this area is always concentrated at moderate $q$.

Resolution of the paradox for the classical case is transparent enough. 
$q_a$ receives commensurable contributions both from large and small $q$.
When one desires to locate the origin of the nonperturbative correction to it, 
it should be minded that distribution $d\sigma/d\sigma_R$ is flat 
(``table-top''), 
just translating as a whole with the increase of the Coulomb parameter (cf. dot-dashed and solid curves in Fig. \ref{fig:q-shift}). 
Thus, for evaluation of the Coulomb parameter dependence, 
essential is only its shift \emph{relative} to the cutoff, 
and \emph{any} of its ends may be considered on equal right.

Physically, it may be more appropriate to attribute the origin of the Coulomb correction to moderate $q$, 
when the imposed cutoff $q_R$ is held fixed, whereas the differential cross section ratio $d\sigma_{cl}/d\sigma_R$ depends on the Coulomb parameter. 
The term ``Coulomb correction'' may then sound obscure. 
At evaluation of an increment of the Coulomb correction at changing $Z_1/v$, 
the necessity to integrate the complicated $q$-dependence of $d\sigma_{cl}/d\sigma_R$ over $q$ can be circumvented (in the classical scattering regime), e.g., 
by presenting it as an integral of $\ln q$ over $d\sigma_{cl}/d\sigma_R$:
\begin{equation}\label{}
-\int_0^{\infty}  \frac{dq}{q} \Delta \frac{d\sigma_{cl}}{d\sigma_R}=\int \Delta \ln q \, d\frac{d\sigma_{cl}}{d\sigma_R}.
\end{equation}
Then, $\Delta \ln q=\Delta\ln\frac{Z_1}{v}=\text{const}$ (at constant $b$ or $\frac{d\sigma_{cl}}{d\sigma_R}$) carries outside of the integral, giving
$\int d\frac{d\sigma_{cl}}{d\sigma_R}=\frac{d\sigma_{cl}}{d\sigma_R}\big|_{q=0}^{q=\infty}=1$, 
and leading back to (\ref{qa-propto-Z1v}). 

But for a theorist, the low-$b$ point of view, 
when the $b$-distribution is regarded as fixed, 
whereas the cutoff [as in (\ref{ln-qa=ln[-]})] or the subtraction term [as in (\ref{ln-qa=ln-alpha+Ca})] as moving, is more beneficial. 
Not only it revives the literal meaning of the term ``Coulomb correction'', 
but more importantly, permits simple extension to multi-center scatterers. 
For scattering in overlapping fields of several scattering centres screened in a complicated way each, 
crucial is only the absence of overlap of their Coulomb singularities, 
where the screening can be neglected.  
That makes it immediately evident that the Coulomb correction is independent of screening at all.

One may ask yet how small can $b_R$ physically be. 
The physical condition is (\ref{chia<<chiR<<chic}), implying
\begin{equation}\label{bR>>}
b_R\gg\frac{2Z_1Ze^2}{p v \chi_c}\approx\frac{Z_1}{\sqrt{\pi n_a l}}.
\end{equation}
At that, for the Rutherford tail to manifest itself in multiple Coulomb scattering, the right-hand side of (\ref{bR>>}) must be much greater than the nuclear radius, which amounts a few fm. 
With $n_a\sim0.06\text{\AA}^{-3}$, that will be fulfilled if $l\ll1$ m. 
For nonrelativistic incident particles, ionization energy losses become significant much earlier. 


The offered demonstration of equivalence of small-$b$ and low-$q$ points of view given above is very simple, but is of limited applicability, 
since it rests significantly on the classical scaling.
In the quantum case $\frac{Z_1Ze^2}{\hbar v}\lesssim1$, 
$d\sigma/d\sigma_R$ depends on 
$\log q$ and $\log \frac{Z_1Z}{\hbar v}$ independently, 
rather than through their sum only,
whereby the Coulomb parameter dependence of the distribution does not reduce to a pure translation in $\log q$ 
(cf. dashed and dot-dashed curves in Fig. \ref{fig:q-shift}). 
In that case, to establish the validity of the small-$b$ point of view, 
a more detailed analysis is needed, which will be carried out in the next section.








\section{2d derivation of Eq. (\ref{q_a-propto-e^f})}\label{sec:2d}

Let us now proceed to computing limit (\ref{qa-lim-int}) for scattering in a screened Coulomb potential under generic conditions 
-- in the absence of central symmetry and for arbitrary Coulomb parameter.
The task is facilitated by the use of techniques of Sec. \ref{subsec:mean-q2}, 
based on representation (\ref{q2-under-a2}), (\ref{qa}) for $q^2 d\sigma/d^2q$.
Inserting
\begin{equation}\label{alpha-def}
\frac{d\sigma_R}{dq}=\frac{8\pi\hbar^2\alpha^2}{q^3}, \qquad \alpha=\frac{Z_1Ze^2}{\hbar v}
\end{equation}
to (\ref{qa-lim-int}), we bring the encountered integral to form
\begin{equation}\label{int-qa2ialpha2}
\int_0^{q_R} \frac{dq}{q}\frac{d\sigma}{d\sigma_R}=\frac{1}{2\pi\hbar^2}\int_{q<q_R} d^2q \left|\frac{\bm q a}{2i\alpha}\right|^2.
\end{equation}

The only caveat is that in order to perform rigorous integration over the \emph{restricted} $\bm q$ plane, it will be necessary first to isolate the Coulomb singularity contribution by breaking $\int d^2b$ in (\ref{qa}) into two parts: 
one over a disk centered at the origin, $b<b_R$, 
and another one over its exterior $b>b_R$. 
In contrast to the previous section, 
the boundary $b_R$ is not assumed to be related to $q_R$, and is chosen so small that at $b<b_R$ the screening is entirely negligible:
\[
\frac{\chi_0(b)}{\hbar}=2\alpha S_0(b), \quad S_0(b)\underset{b<b_R}\simeq \ln\frac{b}{b_0}.
\]
The actual knowledge of $b_0$ will not be needed for us in what follows.
Evaluating the derivative
\[
\frac{\partial}{\partial\bm{b}}e^{2i\alpha\ln\frac{b}{b_0}}=2i\alpha\frac{\bm b}{b^2}\left(\frac{b}{b_0}\right)^{2i\alpha}
\]
and inserting to (\ref{qa-lim-int}), (\ref{int-qa2ialpha2}), (\ref{qa}), we get
\begin{eqnarray}\label{qa=ln-|int+int|^2}
\ln q_a(\alpha)+\frac12=\underset{q_R\to\infty}\lim\Bigg[\ln q_R \qquad\qquad\qquad\qquad\qquad\qquad\nonumber\\
-\frac1{(2\pi)^3\hbar^2}\int_{q<q_R} d^2q\Bigg|\int_{b<b_R}d^2b e^{\frac{i\bm{q}\cdot\bm{b}}{\hbar}}\frac{\bm{b}}{b^2}\left(\frac{b}{b_0}\right)^{2i\alpha}\qquad\qquad\nonumber\\
+\int_{b>b_R}d^2b 
e^{\frac{i\bm{q}\cdot\bm{b}}{\hbar}} e^{2i\alpha S_0(b)} 
\frac{\partial}{\partial\bm{b}} S_0 \Bigg|^2\Bigg].\quad
\end{eqnarray}

When the encountered square of the sum is expanded, the Rutherford singularity remains only in the square of the first term, 
while the interference term vanishes in the limit $b_R\to 0$. 
That leads to representation
\begin{eqnarray}\label{lnqa=int-b<bR+int-b>bR}
\ln q_a(\alpha)+\frac12=\underset{q_R\to\infty}\lim\bigg[\ln q_R
-\mathcal{I}_{\text{hard}}(q_R,b_R,\alpha)\quad\nonumber\\
-\mathcal{I}_{\text{soft}}(q_R,b_R,\alpha)\bigg],
\end{eqnarray}
where
\begin{equation}\label{I1-def}
\mathcal{I}_{\text{hard}}=\frac1{(2\pi)^3\hbar^2}\int_{q<q_R} d^2q\left|\int_{b<b_R}d^2b e^{\frac{i\bm{q}\cdot\bm{b}}{\hbar}}\frac{\bm{b}}{b^2}b^{2i\alpha}\right|^2
\end{equation}
(the constant phase factor $b_0^{-2i\alpha}$ dropped out after squaring),
and
\begin{equation}\label{I2-def}
\mathcal{I}_{\text{soft}}=\frac1{(2\pi)^3\hbar^2}\int_{q<q_R} d^2q\left|\int_{b>b_R}d^2b 
e^{\frac{i\bm{q}\cdot\bm{b}}{\hbar}} e^{2i\alpha S_0(b)} \frac{\partial}{\partial\bm{b}} S_0 \right|^2.
\end{equation}
Integrals $\mathcal{I}_{\text{soft}}$, $\mathcal{I}_{\text{hard}}$ can now be handled independently.

At calculation of $\mathcal{I}_{\text{soft}}$, 
it is already safe to send $q_R\to\infty$, 
because by virtue of the $b>b_R$ cutoff, 
large-$q$ asymptotics of the $b$-integral becomes steeper than in Rutherford's law. 
The integration over the complete $\bm{q}$ plane then proceeds similarly  to Eq. (\ref{ms=msBorn}):
\begin{eqnarray}\label{Isoft=intS2}
\mathcal{I}_{\text{soft}}(b_R)=\int_{b_R}^{\infty}dbb \left| 
 e^{2i\alpha S_0(b)}S'_0\right|^2
\equiv\int_{b_R}^{\infty}dbb S'^2_0(b)\nonumber\\
=\int_{b_R}^{\infty}\frac{db}{b} S_1^2(b).\qquad\qquad\qquad\qquad\qquad
\end{eqnarray}
Most importantly, the eikonal phase factor $e^{2i\alpha S_0(b)}$ has canceled, as in (\ref{ms=msBorn}).
This corroborates the possibility to regard in the impact parameter representation the moderate-$b$ contribution to the Coulomb correction to $q_a$ as absent. 
Since the eikonal factor has canceled in $\mathcal{I}_{\text{soft}}$, 
this contribution is independent of $\alpha$, 
and can be related with $q_a(0)$ by letting $\alpha=0$ in Eq. (\ref{lnqa=int-b<bR+int-b>bR}):
\begin{eqnarray}\label{lim-lim}
\ln q_a(0)+\frac12
=\underset{q_Rb_R\to\infty}\lim\left[\ln q_R b_R
-\mathcal{I}_{\text{hard}}(q_R,b_R,0)\right]\nonumber\\
-\underset{b_R\to0}\lim\left[\mathcal{I}_{\text{soft}}(b_R)+\ln b_R\right].\qquad\qquad
\end{eqnarray}

Evaluation of $\mathcal{I}_{\text{hard}}$ is complicated by the effect of singularity of the integrand, 
but is facilitated by its independence of screening. 
To eliminate $\alpha$-dependence in the integrand of the inner integral in (\ref{I1-def}), we integrate over the azimuth of $\bm{b}$, and rescale the integration variable $b=\hbar x/q$,
\begin{equation*}\label{int-b<bR}
\left|\int_{b<b_R}d^2b e^{\frac{i\bm{q}\cdot\bm{b}}{\hbar}}\frac{\bm{b}}{b^2}b^{2i\alpha}\right|^2
=\left|\frac{2\pi\hbar}{q}\int_{0}^{qb_R/\hbar}dxx^{2i\alpha} J_1\left(x\right) \right|^2.
\end{equation*}
Inserting this to the $q$ integral in (\ref{I1-def}) brings it to form
\begin{equation*}\label{}
\mathcal{I}_{\text{hard}}(q_Rb_R,\alpha)
=\int_0^{q_Rb_R/\hbar} \frac{ds}{s}
\left|\int_{0}^{s}dxx^{2i\alpha} J_1\left(x\right) \right|^2, 
\end{equation*}
\[
s=\frac{qb_R}{\hbar}.
\]
Benefiting from the fact that the inner integral depends on $s$ only through its upper limit, 
$\mathcal{I}_{\text{hard}}$ can be reduced to a double integral by integrating over $s$ by parts:
\begin{eqnarray*}\label{}
\mathcal{I}_{\text{hard}}(q_Rb_R,\alpha)\qquad\qquad\qquad\qquad\qquad\qquad\qquad\nonumber\\
=\ln\frac{q_Rb_R}{\hbar}-\int_0^{\infty} ds\ln s \frac{d}{ds}
\left|\int_{0}^{s}dxx^{2i\alpha} J_1\left(x\right) \right|^2\nonumber\\
\equiv\ln\frac{q_Rb_R}{\hbar}\qquad\qquad\qquad\qquad\qquad\qquad\qquad\quad\,\,\nonumber\\
-2\mathfrak{Re}\int_0^{\infty} ds s^{-2i\alpha}\ln s\,J_1\left(s\right)
\int_{0}^{s}dxx^{2i\alpha} J_1\left(x\right).
\end{eqnarray*}
(In the remaining integral, it was justified to send the upper integration limit $qb_R/\hbar$ to infinity).
Thereby $q_Rb_R$ and $\alpha$ dependencies separate. 
Coulomb correction is entirely contained in the latter. 

The remaining double integral can be simplified by rescaling once again the integration variable: $x=sy$, which eliminates one of the $\alpha$-dependent factors in the integrand:
\begin{eqnarray}\label{ln-I1=intdQdy}
\ln\frac{q_Rb_R}{\hbar}-\mathcal{I}_{\text{hard}}(q_Rb_R,\alpha)\qquad\qquad\qquad\qquad\nonumber\\
=2\mathfrak{Re}\int_0^{\infty} ds s\ln s\,J_1\left(s\right)\int_{0}^{1}dyy^{2i\alpha} J_1\left(sy\right).
\end{eqnarray}
Evaluation of the double integral here is impeded by the fact that inner integral $\int_{0}^{1}dyy^{2i\alpha} J_1\left(sy\right)$ is a complicated function of both $s$ and $\alpha$. 
The difficulty might be circumvented by changing the integration order, but insofar as $s$-integral is improper, this is generally illegitimate 
(and may be the root of misconceptions). 
Formal application of this procedure yields an integral 
\begin{equation}\label{y/(1-y^2)}
\int_0^{\infty} ds s\ln s\,J_1\left(s\right)J_1\left(sy\right)
=-\frac{y}{1-y^2},
\end{equation}
which diverges at $y\to1$, causing in turn a divergence of the $y$-integral on the upper limit, 
whereas physically the result must be finite.

Fortunately, the encountered divergence is suppressed, and the procedure becomes legitimate, when applied to a difference $\mathcal{I}_{\text{hard}}(q_Rb_R,0)-\mathcal{I}_{\text{hard}}(q_Rb_R,\alpha)$ needed for the ratio $q_a/q_a(0)$ [cf. Eq. (\ref{ln-qa=intdsigma-dsigma1})]. 
With the aid of (\ref{y/(1-y^2)}), it evaluates to a finite result:
\begin{eqnarray}\label{I-I=Repsi}
\mathcal{I}_{\text{hard}}(q_Rb_R,0)
-\mathcal{I}_{\text{hard}}(q_Rb_R,\alpha)\qquad\qquad\qquad\qquad \nonumber\\
=2\mathfrak{Re}\int_0^{\infty} ds s\ln s\,J_1\left(s\right)
\int_{0}^{1}dy(y^{2i\alpha}-1) J_1\left(sy\right)\nonumber\\
=2\mathfrak{Re}\int_{0}^{1}dy y  \frac{1-y^{2i\alpha}}{1-y^2}
=\mathfrak{Re}\psi(1+i\alpha)+\gamma_{\text{E}}\qquad\nonumber\\
=f(\alpha).\qquad\qquad\qquad\qquad\qquad\qquad\qquad\qquad\quad\,\,\,
\end{eqnarray}
As for integral $\mathcal{I}_{\text{hard}}(q_Rb_R,0)$, its evaluation can be accomplished without changing the integration order:
\begin{eqnarray}\label{lnq-I=ln2-gammaE}
\ln\frac{q_Rb_R}{\hbar}-\mathcal{I}_{\text{hard}}(q_Rb_R,0)\qquad\qquad\qquad\qquad\nonumber\\
=2\int_0^{\infty} ds s\ln s\,J_1\left(s\right)\int_{0}^{1}dy J_1\left(sy\right)\qquad\quad\,\,\,\nonumber\\
=2\int_0^{\infty} ds \ln s\,J_1\left(s\right)\left[1- J_0\left(s\right)\right]
=\ln 2-\gamma_{\text{E}}.
\end{eqnarray}
Subtracting Eq. (\ref{ln-I1=intdQdy}) from (\ref{lnqa=int-b<bR+int-b>bR}) and employing Eq. (\ref{I-I=Repsi}), we recover Eq. (\ref{q_a-propto-e^f}), while with the aid of Eqs. (\ref{lnq-I=ln2-gammaE}) and (\ref{Isoft=intS2}) $q_a(0)$ checks to satisfy Eq. (\ref{qa-lim-bto0}).

The presented derivation based on partial integration and change of the integration order is general and sufficiently natural (not resorting to any artificial regulating functions).
It demonstrates that although in the quantum case the shape of the $b$-distribution becomes Coulomb parameter dependent, in its $q^2$-weighted integral this dependence drops out, as in Sec. \ref{subsec:mean-q2}. 
Granted that at not too small impact parameters eikonal phases exactly cancel after the integration over all $q$, 
there is no difference of the ``soft'' contribution from its Born approximation, 
and this needs not be incorporated in the theory manually, 
based on physical arguments. 
In the impact parameter representation of high-energy quantum mechanics, 
Coulomb correction may generally be regarded as stemming from small $b$, 
where the integration interval is restricted by a nonzero boundary value, 
making the cancellation incomplete. 
Since in quantum mechanics a sharp cutoff in $q$ does not correspond to a sharp cutoff in $b$, the dependence on the Coulomb parameter becomes more sophisticated than in classical mechanics (see Sec. \ref{sec:prelim}), being expressed by function (\ref{f-def}). 

Having clarified the issue of the physical and mathematical origin of Coulomb corrections, we are now in a position to apply our technique to a more realistic model of the atom.  




\section{Inelastic scattering}\label{sec:inel-scat}

Microscopically, every atom, especially at low $Z$, must be treated as a few-body system, containing a finite number of electrons, rather than just a field of the nucleus screened by the mean electron charge distribution. 
For simplicity, let the incident particle be structureless. 
It can experience close collisions with the atomic nucleus and atomic electrons. 
Note that in perturbation theory, hard collisions with electrons are equivalent to inelastic, 
but once we go beyond perturbative treatment, 
close encounter with the nucleus does not preclude possibility of atom ionization by soft interactions with atomic electrons, etc. 
For description of multiple Coulomb scattering, though, we need only the inclusive differential cross section and its Rutherford asymptotics, 
rather than separation into elastic and inelastic channels.

\subsection{Inclusive Moli\`{e}re angle}

During a fast collision of the incident high-energy particle with an atom, specifically, under conditions \cite{fast-collision-conditions}
\begin{equation}\label{frozen-condition}
v\gg v_a,\, v_{a1},
\end{equation}
where $v_a\sim \frac{Z^{1/3}e^2}{\hbar}$ and $v_{a1}\sim \frac{Z_1^{1/3}e^2}{\hbar}$ are typical values of velocities of electrons bound at the target atom and the incident ion (in case of electron capture), 
all the atomic electrons may be regarded as static. 
Their coordinate distribution is determined by the initial-state wave function.  
Static pointlike electrons (sitting at positions $\bm r_1,\ldots, \bm r_Z$) and the nucleus (located at the origin) create a classical electrostatic 
(though not spherically symmetric) field, 
the amplitude of scattering of the projectile in which can still be evaluated in the eikonal approximation:
\begin{eqnarray}\label{ar1Z}
a(\bm{r}_1,\ldots,\bm{r}_Z;\bm{q})=\frac{1}{2\pi i\hbar}\int d^2 b e^{i\bm{q}\cdot\bm{b}/\hbar}\qquad\qquad\quad\nonumber\\
\times\left\{1-e^{\frac{i}{\hbar}\left[\chi_{0A}(b)+\sum_{k=1}^{Z}\chi_{0e}(|\bm{b}-\bm{b}_k|)\right]}\right\}.
\end{eqnarray}
Function $\chi_0$ is now explicitly known:
\begin{equation}\label{chi0=sumAe}
\frac{1}{\hbar}\chi_0(\bm b_1,\ldots,\bm Z;\bm b)
=\frac{Z_1Ze^2}{\hbar v}\ln b
-\frac{Z_1e^2}{\hbar v}\sum_{k=1}^Z\ln|\bm b-\bm b_k|.
\end{equation}

To convert (\ref{ar1Z}) to an amplitude of scattering with excitation or ionization of the atom from initial (normally, ground) state $|0\rangle$ to arbitrary (discrete or continuum) state $|n\rangle$, 
(\ref{ar1Z}) must be convolved with wave functions $\psi_0$, $\psi_n$ of those states \cite{Glauber}:
\begin{eqnarray}\label{na0-def}
\langle n|a|0\rangle=\sum_{s_1,\ldots,s_Z}\int d^3r_1\ldots d^3r_Z \psi_n^*(\bm{r}_1,s_1,\ldots,\bm{r}_Z,s_Z) \nonumber\\
\times a(\bm{r}_1,\ldots,\bm{r}_Z;\bm{q})
\psi_0(\bm{r}_1,s_1,\ldots,\bm{r}_Z,s_Z).\quad
\end{eqnarray}
($s_1,\ldots,s_Z$ are electron spin quantum numbers.)

Next, amplitudes (\ref{na0-def}) are squared to yield partial differential cross sections,
and summed over all $n$ to give the inclusive differential cross section
\begin{eqnarray}\label{dsigma0t-def}
\frac{d\sigma_{0}}{d^2q}=
\sum_{n=0}^{\infty}\left|\left\langle n|a|0\right\rangle\right|^2 
=\langle 0|a^* a|0\rangle\qquad\qquad\qquad\qquad\nonumber\\
\equiv\int d^3r_1\ldots d^3r_Z \rho_0(\bm{r}_1,\ldots,\bm{r}_Z)
\left|a(\bm{r}_1,\ldots,\bm{r}_Z;\bm{q})\right|^2.
\end{eqnarray}
Here 
\begin{eqnarray}
\rho_0(\bm{r}_1,\ldots,\bm{r}_Z)
=\sum_{s_1,\ldots,s_Z}|0\rangle\langle 0|\qquad\qquad\qquad\nonumber\\
=\sum_{s_1,\ldots,s_Z}\left|\psi_0(\bm{r}_1,s_1,\ldots,\bm{r}_Z,s_Z)\right|^2
\end{eqnarray} 
is the normalized initial-state electron coordinate distribution,
\[
\int d^3r_1\ldots d^3r_Z \rho_0(\bm{r}_1,\ldots,\bm{r}_Z)=1,
\]
and subscript $0$ at $d\sigma$ indicates that the initial state was 
$|0\rangle$.

Ultimately, (\ref{dsigma0t-def}) is inserted in the definition of $q_a$ similar to (\ref{qa-lim-int}):
\begin{equation}\label{qat-def}
\ln q_{at}+\frac12=\underset{q_R\to\infty}\lim\left(\ln q_R-\int_0^{q_R} \frac{dq}{q}\frac{d\sigma_0}{d\sigma_R}\right),
\end{equation}
where $d\sigma_R$ now is high-$q$ asymptotics of $d\sigma_0$:
\begin{equation}\label{}
\frac{d\sigma_R}{dq}=\frac{2\pi Z(Z+1)}{q^3}\left(\frac{2Z_1e^2}{v}\right)^2,
\end{equation}
with factor $Z(Z+1)$ taking into account hard scattering on electrons. 
Subsctipt $at$ in (\ref{qat-def}) distinguishes it from $q_a$ in 
Eq. (\ref{qa-lim-int}), 
which is conventionally reserved for elastic cross section 
\cite{Fano,Bond-correl}.

\subsection{Isolation of the Coulomb correction}

Computation of the averaged differential cross section (\ref{dsigma0t-def}) is more complicated than that for scattering in the averaged atomic potential, 
but the procedure of isolation of the Coulomb correction remains largely the same. 
The key point is that integration over the electron coordinates and summation over final states can be deferred to the last stage of the calculation, 
by interchanging the limiting procedure and the initial state averaging:
\begin{eqnarray}\label{interchange-the-order}
\ln q_{at}+\frac12
=\int d^3r_1\ldots d^3r_Z \rho_0(\bm{r}_1,\ldots,\bm{r}_Z)\qquad\qquad\qquad\nonumber\\
\times\underset{q_R\to\infty}\lim\left[\ln q_R
-\frac{1}{2\pi Z(Z+1)}\left(\frac{v}{2Z_1e^2}\right)^2
\! \int_{q<q_R} \! d^2q q^2|a|^2\right]\nonumber\\
=\int d^3r_1\ldots d^3r_Z \rho_0(\bm{r}_1,\ldots,\bm{r}_Z)\qquad\qquad\qquad\nonumber\\
\times\underset{q_R\to\infty}\lim\Bigg[\ln q_R
-\frac{1}{(2\pi)^3 Z(Z+1)}\left(\frac{v}{2Z_1e^2}\right)^2\quad\nonumber\\
\times\int_{q<q_R} d^2q \left|\int d^2b
e^{\frac{i}{\hbar}\bm{q}\cdot\bm{b}}\frac{\partial}{\partial\bm{b}}e^{\frac{i}{\hbar}\chi_0(\bm{b})}\right|^2\Bigg],\qquad
\end{eqnarray}
where in the second equality we employed Eq. (\ref{qa}), 
and $\chi_0$ is given by Eq. (\ref{chi0=sumAe}).
The last line in (\ref{interchange-the-order}) involves a multiple integral, which can be evaluated similarly to Sec. \ref{sec:2d}, provided we isolate \emph{all} the singularities in the Coulomb $\bm b$ plane:
\[
\int d^2b=\int_{b<b_R} d^2b+\sum_{k=1}^Z \int_{|\bm{b}-\bm{b}_k|<b_R} d^2b
+\int_{b,|\bm{b}-\bm{b}_k|>b_R} d^2b.
\]
Once the square of this sum of $\bm b$-integrals is expanded, 
and limit $b_R\to0$ is taken, 
the interference between the partial $\bm b$-integrals vanishes, as in Sec. \ref{sec:2d}:
\begin{eqnarray}\label{limIhard+tildeIsoft}
\ln q_{at}+\frac12
=\underset{q_R\to\infty}\lim
\Bigg[\ln q_R
-\frac{Z}{Z+1}\mathcal{I}_{\text{hard}}\left(q_Rb_R,\frac{Z_1Ze^2}{\hbar v}\right)\nonumber\\
-\frac{1}{Z+1}\mathcal{I}_{\text{hard}}\left(q_Rb_R,\frac{Z_1e^2}{\hbar v}\right)\nonumber\\
-\tilde{\mathcal{I}}_{\text{soft}}\left(q_R,b_R,\frac{Z_1Ze^2}{\hbar v},\frac{Z_1e^2}{\hbar v}\right)
\Bigg].
\end{eqnarray}
Here $\mathcal{I}_{\text{hard}}\left(q_Rb_R,\alpha\right)$ is given by Eq. (\ref{I1-def}) (since we neglect overlaps of the regions contributing to the Coulomb correction, they are independent of the electron distribution at all), and 
\begin{eqnarray}\label{tildeIsoft-def}
\tilde{\mathcal{I}}_{\text{soft}}=\int d^3r_1\ldots d^3r_Z \rho_0(\bm{r}_1,\ldots,\bm{r}_Z)\qquad\qquad\quad\nonumber\\
\times
\frac{1}{(2\pi)^3 Z(Z+1)}\left(\frac{v}{2Z_1e^2}\right)^2\qquad\qquad\qquad\quad\nonumber\\
\times\int_{q<q_R} d^2q \left|\int_{b,|\bm{b}-\bm{b}_k|>b_R} d^2b
e^{\frac{i}{\hbar}\bm{q}\cdot\bm{b}}\frac{\partial}{\partial\bm{b}}e^{\frac{i}{\hbar}\chi_0(\bm{b})}\right|^2.
\end{eqnarray}

In $\tilde{\mathcal{I}}_{\text{soft}}$, similarly to Sec. \ref{sec:2d}, it is legitimate to send $q_R\to\infty$, whereupon closure identity applies, and eikonal phases cancel:
\begin{eqnarray}\label{}
\tilde{\mathcal{I}}_{\text{soft}}=\tilde{\mathcal{I}}_{\text{soft}}(b_R,Z)
=\int d^3r_1\ldots d^3r_Z \rho_0(\bm{r}_1,\ldots,\bm{r}_Z)\quad\nonumber\\
\times
\frac{1}{2\pi Z(Z+1)}\left(\frac{v}{2Z_1e^2}\right)^2
\int_{b,|\bm{b}-\bm{b}_k|>b_R} \! d^2b
 \left[ \frac{\partial}{\partial\bm{b}}\chi_0(\bm{b})\right]^2.
\end{eqnarray}
Since it becomes independent of the Coulomb parameter, 
similarly to Eq. (\ref{lim-lim}), we can relate it to $q_{at}(0)$:
\begin{equation}\label{}
\tilde{\mathcal{I}}_{\text{soft}}\left(b_R\right)
\! =\underset{q_Rb_R\to\infty}\lim \!
\left[\ln q_R
-\mathcal{I}_{\text{hard}}\left(q_Rb_R,0\right)\right]
-\ln q_{at}(0)-\frac12.
\end{equation}
Eq. (\ref{limIhard+tildeIsoft}) can then be expressed as
\begin{eqnarray}\label{}
\ln q_{at}+\frac12
=\ln q_{at}(0)\qquad\qquad\qquad\qquad\qquad\qquad\nonumber\\
+\frac{Z}{Z+1}
\left[\mathcal{I}_{\text{hard}}\left(q_Rb_R,0\right)
-\mathcal{I}_{\text{hard}}\left(q_Rb_R,\frac{Z_1Ze^2}{\hbar v}\right)\right]\nonumber\\
+\frac{1}{Z+1}\left[\mathcal{I}_{\text{hard}}\left(q_Rb_R,0\right)
-\mathcal{I}_{\text{hard}}\left(q_Rb_R,\frac{Z_1e^2}{\hbar v}\right)\right],
\end{eqnarray}
or, employing Eq. (\ref{I-I=Repsi}),
\begin{equation}\label{expff}
\frac{q_{at}}{q_{at}(0)}=e^{\frac{Z}{Z+1}f\left(ZZ_1e^2/\hbar v\right)+\frac{1}{Z+1}f\left(Z_1e^2/\hbar v\right)}.
\end{equation}
It can be straightforwardly generalized to scatterers consisting of arbitrary charges $Z_k e$ (electrons and nuclei of atoms of different sorts):
\begin{equation}\label{}
\frac{q_{at}}{q_{at}(0)}=\exp\left[\frac{\sum_k Z_k^2 f\left(\frac{Z_kZ_1 e^2}{\hbar v}\right)}{\sum_k Z_k^2}\right].
\end{equation}
This structure has the same appearance as for multiple Coulomb scattering in a composite substance \cite{Moliere,Scott}.

Eq. (\ref{expff}), which is the main result of this article, 
thus has a transparent meaning. 
Since Coulomb correction is independent of screening, 
contributions to it from atomic electrons and nuclei may be calculated independently, 
like stemming from scattering on bare particles in an ideal plasma.
A word of caution must be sound yet that electronic contribution $\frac{1}{Z+1}f\left(Z_1e^2/\hbar v\right)$ is neither purely inelastic nor elastic. 
Only at the Born level it holds that 
\begin{equation}\label{}
q_{at}(0)=q_{a}^{\frac{Z}{Z+1}}(0)q_{in}^{\frac{1}{Z+1}}(0).
\end{equation}
In the Fano theory \cite{Fano}, term containing $f\left(Z_1e^2/\hbar v\right)$ [essentially the Bloch correction, cf. Eq. (\ref{Bloch})] is absent, 
whereas $f\left(ZZ_1e^2/\hbar v\right)$, 
approximated as in (\ref{Moliere-sqrt}), 
without $\frac{Z}{Z+1}$ factor, is attributed to the elastic channel.
Thus, \cite{Fano}  takes into account the inelastic contribution to $q_{at}(0)$, but not to its Coulomb correction. 
As we had pointed out at the beginning of Sec. \ref{sec:inel-scat}, 
\[
q_a\neq e^{\frac{Z}{Z+1}f\left(ZZ_1e^2/\hbar v\right)}q_a(0),
\]
\[
q_{in}\neq e^{\frac{1}{Z+1}f\left(Z_1e^2/\hbar v\right)}q_{in}(0),
\]
because evolution of $q_a$ and $q_{in}$ is not independent (they intermix).
Nonetheless, in \emph{inclusive} calculation any processes (elastic or inelastic) taking place before and after the particle approach to the Coulomb singularity are irrelevant, whence the ``strength'' (partial contribution) of each singularity is determined only by its charge, exactly resembling the situation in a composite substance.

\begin{figure}
\includegraphics{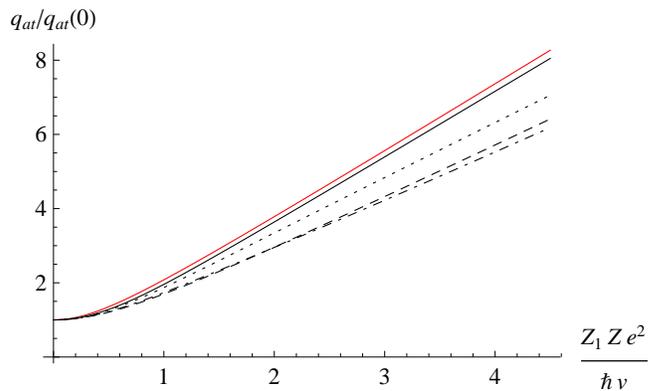}
 \caption{\label{fig:Zf+f} 
Ratio (\ref{expff}) of the Moli\`{e}re screening angle to its Born value 
[$\chi_{at}/\chi_{at}(0)=q_{at}/q_{at}(0)$] at $ZZ_1e^2/\hbar v\to0$.
Black solid curve, for $Z=1$, or for $Z>50$. 
Dashed, $Z=2$ (helium). 
Dot-dashed, $Z=4$ (beryllium). 
Dotted, $Z=14$ (silicon). 
Red solid (the uppermost) curve, Moli\`{e}re's algebraic interpolation (\ref{Moliere-sqrt}).
}
\end{figure}

\subsection{Analysis}

The obtained structure (\ref{expff}), unlike (\ref{q_a-propto-e^f}), depends on two variables.
Only for $Z\gg1$ or $Z=1$ (as long as for hydrogen atom, 
high-energy Rutherford scattering at its nucleus and electron is largely equivalent), 
its exponent reduces to $f\left(ZZ_1e^2/\hbar v\right)$, 
leading back to Eq. (\ref{q_a-propto-e^f}). 
Thus, incoherent nonperturbative effects manifest themselves for intermediate $Z$. 
To illustrate the deviation of (\ref{expff}) from Eq. (\ref{q_a-propto-e^f}), 
in capacity of its two independent variables it is convenient to take $Z$ and $ZZ_1e^2/\hbar v$. 
Fig. \ref{fig:Zf+f} shows that for any $Z$, 
ratio (\ref{expff}) monotonously (asymptotically -- linearly) increases with $ZZ_1e^2/\hbar v$, but its slope is $Z$-dependent. 
To quantitatively characterize this dependence, 
it may suffice to consider two extreme cases.

\begin{figure}
\includegraphics{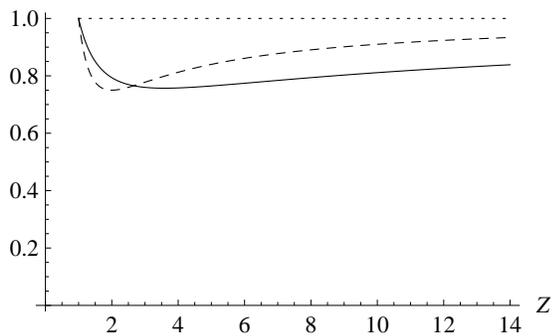}
 \caption{\label{fig:1-075-1} 
$Z$-dependence of factors determining deviation from the pure coherent contribution to $q_a$, for large and for small Coulomb parameters. 
Solid curve, factor $Z^{-\frac{1}{Z+1}}$ entering Eq. (\ref{Z-1Z+1*coher}). 
Dashed, factor $1-Z^{-1}+Z^{-2}$ entering Eq. (\ref{1-Z-1+Z-2*coher}). 
The relative departure of both factors from 1 is $\lesssim25\%$.
}
\end{figure}

At $ZZ_1e^2/\hbar v<1$ (e.g., for relativistic incident electrons or protons, 
when $Z_1e^2/\hbar v\sim 10^{-2}$, and $Z<10^2$ for all substances), 
arguments of both functions $f$ in (\ref{expff}) are small. 
Then this function may be expanded to the leading, quadratic order:
\[
f(s)\underset{s\to0}\simeq\zeta(3)s^2,
\]
where $\zeta(3)=-\psi''(1)/2=\sum_{n=1}^{\infty}n^{-3}=1.202$, 
yielding
\begin{eqnarray}\label{1-Z-1+Z-2*coher}
\ln\frac{q_{at}}{q_{at}(0)}
&=&\zeta(3)\frac{Z^3+1}{Z+1}\left(\frac{Z_1 e^2}{\hbar v}\right)^2\nonumber\\
&\equiv&\left(1-Z^{-1}+Z^{-2}\right)\zeta(3)\left(\frac{ZZ_1 e^2}{\hbar v}\right)^2.
\end{eqnarray}
Factor $1-Z^{-1}+Z^{-2}$ multiplying the pure coherent result $\zeta(3)\left(\frac{ZZ_1 e^2}{\hbar v}\right)^2$ corresponding to Eq. (\ref{q_a-propto-e^f}) has a minimum at $Z=2$, where it equals $3/4$, whereas at $Z=1$ and $Z\to\infty$ it turns to unity (see Fig. \ref{fig:1-075-1}, dashed curve).

In the opposite case $Z_1e^2/\hbar v>1$ (e.g., for $\alpha$-particles with energies $T<500$ keV) arguments of both functions $f$ in (\ref{expff}) are large.
Then one can employ the large-argument asymptotics (\ref{fs->infty}) to get
\begin{equation}\label{Z-1Z+1*coher}
\frac{q_{at}}{q_{at}(0)}\simeq Z^{-\frac{1}{Z+1}} \frac{Z_1Z e^2}{\hbar v}  e^{\gamma_{\text{E}}}.
\end{equation}
The last two factors here may be identified with the coherent contribution 
[cf. Eq. (\ref{qa-classical-limit})], 
whereas the first factor $Z^{-\frac{1}{Z+1}}$ has a minimum at $Z=3.6$, where it equals $0.756$, 
while turning to unity at $Z=1$ and $Z\to\infty$ 
(see Fig. \ref{fig:1-075-1}, solid curve). 
Hence, at $ZZ_1e^2/\hbar v\gtrsim1$ and low $Z\neq1$, 
the (negative) difference of $q_{at}/q_{at}(0)$ from its coherent counterpart (\ref{q_a-propto-e^f}) or (\ref{qa-classical-limit}) is $\sim25\%$,  
remaining sizable also at medium $Z$ (see the dotted curve for silicon in Fig. \ref{fig:Zf+f}).

\section{Practical remarks}\label{sec:pract}

At low $Z$, the found difference of the Coulomb correction from the pure coherent scattering approximation (\ref{Moliere-sqrt}) or (\ref{q_a-propto-e^f}) is much more significant than the inaccuracy \cite{Nigam-Sundaresan-Wu} of the Moli\`{e}re parametrization (the difference between the two solid curves in Fig. \ref{fig:Zf+f}).
But Moli\`{e}re theory was tested against a broad variety of target materials, projectile types and energies, 
always with a satisfactory agreement, 
so one may be surprised why deviations from it remained unnoticed so far.

In this regard, it should be minded first that in the Moli\`{e}re theory $\chi_a$ enters only as a constant correction to the large logarithm $\ln\frac{\chi_c^2}{\chi_a^2}$ (see \cite{Moliere,Scott} and Appendix \ref{app:how-qa-arises}). 
Since there are no observables exceptionally sensitive to $\chi_a$, 
the most practical among them is that, for which the statistics is the highest, i.e., the width of the angular distribution. 
The applicability of Moli\`{e}re's multiple scattering theory demands $\frac{\chi_c^2}{\chi_a^2}>10^2$ \cite{Scott,Bond-correl}, 
wherewith $\ln\frac{\chi_c^2}{\chi_a^2}>5$. 
Since, as we determined at the end of Sec. \ref{sec:inel-scat}, 
the deviation from Moli\`{e}re's prediction for the screening angle $\chi_a$ will not exceed 25\%, 
relative effects in the distribution width stemming from the electronic contribution to $q_a$ numerically do not exceed
$\frac{\Delta\chi_a}{\chi_a \ln{\chi_c^2}/{\chi_a^2}}\lesssim0.25/5=0.05$. 
Such minor corrections may elude attention for a long time, 
but should ultimately become manifest in high-statistics experiments. 

Since the sensitivity to $\chi_a$ is not high (see also \cite{Meyer-Krygel}), it is important at least to determine optimal conditions for its measurement.
To ensure that incident charged particles are pointlike (fully stripped ions only), in practice it is simplest to use proton beams ($Z_1=1$). 
To make $q_{at}/q_{at}(0)$ substantially different from unity, 
one needs to provide a sizable Coulomb parameter, 
but too low velocities are undesirable, 
insofar as energy losses become commensurable with the particle energy.  
Relative deviations close to maximal $\Delta q_{at}/q_{at}(0)\sim 0.25$ can be reached already at 
\begin{equation}\label{Coul-par-sim2}
Z_1Ze^2/\hbar v=Ze^2/\hbar v \sim 2
\end{equation}
(see Fig. \ref{fig:Zf+f}). 
That is still compatible with condition (\ref{frozen-condition}) for the impulse approximation.
Condition (\ref{Coul-par-sim2}) at $Z=6$ (carbon, being one of the most widely used low-$Z$ targets)\footnote{The need for nuclear form factors then does not arise.} 
corresponds to velocities ${v}/{c}\sim \frac{Z}{2*137}=2\cdot 10^{-2}$, 
i.e., to proton energies\footnote{At that, longitudinal momentum $p=\sqrt{2m_p T}\sim 20$ MeV/c is much larger than transverse momentum transfers in Fig. \ref{fig:q-shift}, 
ensuring the validity of the small-angle approximation.} 
$T=m_p v^2/2\sim 200$ keV.

Experiments on multiple scattering of sub-MeV protons on carbon foils were carried out since 1960-ies \cite{Bednyakov-JETP-1962,Bernhard-Krygel-1972}.
Notably, experiment \cite{Bernhard-Krygel-1972} aimed to measure effective screening radii for $T=60$--270 keV protons scattered by carbon and germanium foils, 
and reported excess of the measured screening radii over theoretical predictions (obtained in a $Z_1$-modified Thomas-Fermi approximation \cite{LNS}) by 20\% for carbon and 10\% for germanium. 
That may signal in favor of our predictions,
but theoretical ($\sim10\%$ shell corrections) and experimental (15\% for carbon and 10\% for germanium) errors for this quantity were significant. 
More accurate experiments, including targets of low-$Z$ elements other than carbon, are desirable.


\section*{Acknowledgements}

This work was supported in part by the National Academy of Sciences of Ukraine
(projects 0118U006496, 0118U100327, and 0121U111556).

\appendix

\section{Multiple Coulomb scattering}\label{app:how-qa-arises}


In the multiple Coulomb scattering theory, 
the distribution function $f(\theta,l)$ of particles deflected to a small angle $\theta$ after a path length $l$ obeys transport equation
\begin{equation}\label{transport-eq}
\frac{\partial f}{\partial l}= n_a\int
d\sigma(\chi)\left[f(\bm \theta-\bm \chi,l)-f(\theta,l)\right].
\end{equation}
Here  $d\sigma(\chi)=d^2\chi\frac{d\sigma}{d^2\chi}$ is the
differential cross-section of particle scattering on one atom
through angle $\chi$, 
and $n_a$ is the density of atoms in the amorphous medium. 
Equation (\ref{transport-eq}) with initial condition
\[
f(\bm\theta,0)=\delta(\bm\theta)
\]
admits a generic solution in integral form \cite{Bothe}
\begin{equation}\label{Bothe-solution}
f(\theta,l)=\frac{1}{2\pi}\int_0^{\infty} d\rho\rho J_0(\rho\theta)
e^{-n_al\int d\sigma(\chi) \left[1-J_0(\rho\chi)\right]}.
\end{equation}
Under conditions of multiple scattering, the detailed knowledge of  $d\sigma(\chi)$ is unnecessary.

Since atomic sizes are very small ($n_a\sigma l$ is large), for most of the macroscopic solid targets there are many atoms encountered along the particle path, i.e., the scattering process may be regarded as multiple. 
Then integral (\ref{Bothe-solution}) simplifies. 
With the account of the Coulomb character of atomic scattering, it has to be evaluated with the NLL accuracy.
This proceeds by breaking the integration $\chi$-interval in (\ref{Bothe-solution}) by $\chi_R$ such that 
\begin{equation}\label{chia<<chiR<<chic}
\chi_a\ll\chi_R\ll\chi_c.
\end{equation}

Using the Rutherford asymptotics, and the value of the integral
\[
\int_{\chi_R}^{\infty}\frac{d\chi}{\chi^3}
\left[1-J_0(\rho\chi)\right]=\frac{\rho^2}{4}
\left(\ln\frac{2}{\rho\chi_R}+1-\gamma_{\text{E}}\right),
\]
for the hard part we get
\begin{equation}\label{int-hard-part}
n_a l \int_{\chi_R}^{\infty}d\sigma(\chi)\left[1-J_0(\rho\chi)\right]
=\frac{\chi_c^2\rho^2}{2}
\left(\ln\frac{2}{\rho\chi_R}+1-\gamma_{\text{E}}\right),
\end{equation}
where
\[
\chi_c^2=\frac{4\pi n_a l Z(Z+1)e^4}{p^2 v^2}.
\]

The soft part, with the use of approximation
$1-J_0(\rho\chi)\simeq\frac{1}{4}\rho^2\chi^2$,
\[
n_a l \int_0^{\chi_R}d\sigma(\chi)\left[1-J_0(\rho\chi)\right]
\simeq\frac{\rho^2}{4}n_a l \int_0^{\chi_R}d\sigma(\chi)\chi^2
\]
and definition (\ref{qa-lim-int})
\begin{equation}
\int_0^{\chi_R}\frac{d\chi}{\chi}\frac{d\sigma}{d\sigma_R}\simeq\ln\frac{\chi_R}{\chi_a}-\frac12,
\end{equation}
[which becomes an identity if 
$\frac{d\sigma}{d\sigma_R}=(1+\chi_a^2/\chi^2)^{-2}$], equals
\begin{equation}\label{int-soft-part}
n_a l \int_0^{\chi_R}d\sigma(\chi)\left[1-J_0(\rho\chi)\right]
=\frac{\chi_c^2\rho^2}{2}
\left(\ln\frac{\chi_R}{\chi_a}-\frac12\right).
\end{equation}
Adding up (\ref{int-hard-part}) and (\ref{int-soft-part}), 
we obtain the reduced expression for the exponent
\begin{eqnarray}\label{}
n_a l \int_{0}^{\infty}d\sigma(\chi)\left[1-J_0(\rho\chi)\right]
&=&\frac{\chi_c^2\rho^2}{2}
\left(\ln\frac{2}{\rho\chi_a}+\frac12-\gamma_{\text{E}}\right)\nonumber\\
&\equiv&\frac{\chi_c^2\rho^2}{2}
\ln\frac{2}{\rho\chi'_a},
\end{eqnarray}
which enters the integral representation of the distribution function
\begin{equation}\label{Moliere-approximation}
f(\theta,l)=\frac{1}{2\pi}\int_0^{\infty} d\rho\rho J_0(\rho\theta)
e^{-\frac{\chi_c^2\rho^2}2\ln\frac{2}{\chi'_a\rho}}.
\end{equation}
The only characteristic of the medium entering here is $\chi'_a$ defined by Eq. (\ref{chi'a-def}).



\end{document}